\documentclass[preprint,superscriptaddress,notitlepage,endfloat,titlesec]{revtex4-1}
\usepackage{graphicx,bm,mathtools,color}

\begin{document}

\title{Possible Weyl fermions in the magnetic Kondo system CeSb}

\author{C. Y. Guo}
\affiliation{Center for Correlated Matter and Department of Physics, Zhejiang University, Hangzhou 310058, China}
\author{C. Cao}
\affiliation{Department of Physics, Hangzhou Normal University, Hangzhou 310036, China}
\author{M. Smidman}
\affiliation{Center for Correlated Matter and Department of Physics, Zhejiang University, Hangzhou 310058, China}
\author{F. Wu}
\affiliation{Center for Correlated Matter and Department of Physics, Zhejiang University, Hangzhou 310058, China}
\author{Y. J. Zhang}
\affiliation{Center for Correlated Matter and Department of Physics, Zhejiang University, Hangzhou 310058, China}
\author{F. Steglich}
\affiliation{Center for Correlated Matter and Department of Physics, Zhejiang University, Hangzhou 310058, China}
\affiliation{Max Planck Institute for Chemical Physics of Solids, 01187 Dresden, Germany}
\author{F. C. Zhang}
\affiliation{Center for Correlated Matter and Department of Physics, Zhejiang University, Hangzhou 310058, China}
\affiliation{Collaborative Innovation Center of Advanced Microstructures, Nanjing 210093, China}
\author{H. Q. Yuan*}
\affiliation{Center for Correlated Matter and Department of Physics, Zhejiang University, Hangzhou 310058, China}
\affiliation{Collaborative Innovation Center of Advanced Microstructures, Nanjing 210093, China}

\date{\today}
\begin{abstract}
\textbf{Materials where the electronic bands have unusual topologies allow for the realization of novel physics and have a wide range of potential applications. When two electronic bands with linear dispersions intersect at a point, the excitations could be described as Weyl fermions which are massless particles with a particular chirality.  Here we report evidence for the presence of Weyl fermions in the ferromagnetic state of the low-carrier density, strongly correlated Kondo lattice system CeSb, from electronic structure calculations and angle-dependent magnetoresistance measurements. When the applied magnetic field is parallel to the electric current, a pronounced negative magnetoresistance is observed within the ferromagnetic state, which is destroyed upon slightly rotating the field away.  These results give evidence for CeSb belonging to a new class of Kondo lattice materials with Weyl fermions in the ferromagnetic state.}
\end{abstract}
\maketitle

Topological materials have been found to demonstrate a variety of novel phenomena. For instance, topological insulators are fully gapped in the bulk but display a band inversion leading to distinct behaviour from simple band insulators, such as gapless conducting edge states  \cite{TopoRev,RMPTopo}. More recently gapless topological systems such as Dirac semimetals have also been discovered, where the crystal symmetry prevents the opening of a gap at a point where the bands cross linearly, much like a three-dimensional analogue of graphene \cite{Dirac1,Dirac2,DiracSemiNa3Bi}. When either time reversal or inversion symmetry is broken,  such a Dirac point can be split into a pair of Weyl points, near which the states are well described by Weyl fermions \cite{Weyl1,Weyl2,Weyl3,huang2015weyl}. In addition to being massless, Weyl fermions also have a chirality (either left or right handed) and while a Dirac point requires protection from the crystal symmetry to avoid a gap opening, an isolated Weyl point is topologically protected.

\begin{figure}[h]
\begin{center}
 \includegraphics[width=0.75\columnwidth]{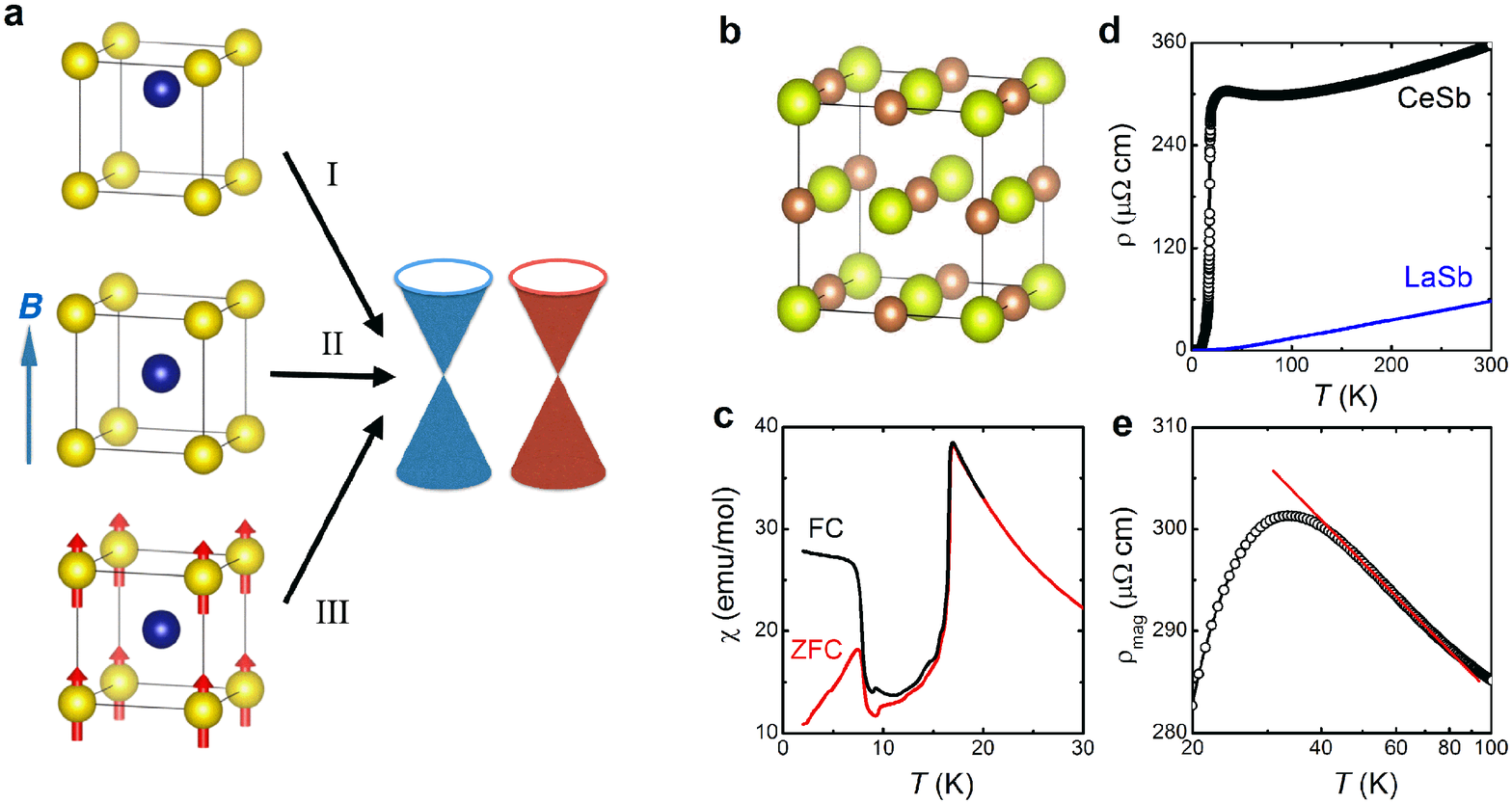}
\end{center}
	\caption{\textbf{(a) Possible routes for achieving Weyl fermions. (b)-(e) Crystal structure and characterization of CeSb.} (a) Schematic diagram for three means of achieving Weyl fermions. Route I is via the breaking of inversion symmetry, route II is via breaking time reversal symmetry by applying a magnetic field, and route III is via breaking time reversal symmetry in the magnetic state. (b) Crystal structure of CeSb. (c) Dc magnetic susceptibility of CeSb in an applied field of 0.1~T measured upon warming after zero-field cooling (ZFC) and field cooling (FC) showing a magnetic transition at around 17~K, and another one around 8~K to the antiferro-ferromagnetic (AFF) phase where the ZFC and FC curves split. (d) Temperature dependence of the electrical resistivity of CeSb and LaSb. (e) Magnetic contribution to the resistivity of CeSb obtained by subtracting the data for LaSb, plotted on a logarithmic temperature scale. The straight line indicates the logarithmic increase of the resistivity with decreasing temperature below about 80~K, suggesting the presence of Kondo scattering.}
   \label{Fig1}
\end{figure}

The realization of Weyl fermions therefore requires either the breaking of inversion or time reversal symmetry. While initially there were proposals for Weyl fermions in some time reversal symmetry breaking magnetic materials  \cite{Weyl1,Weyl2},  experimental evidence was found in materials where inversion symmetry is broken, such as TaAs or WTe$_2$ (route I in Fig.~\ref{Fig1}(a)) \cite{WeylTaAsArc2,WeylTaAsArc3,WeylTaAsArc1,WeylTaAsChirAnom1,NbPChiralAnom,WTe2ChiralAnom}. A second route for creating Weyl points is for time reversal symmetry to be broken by an applied magnetic field (route II in Fig.~\ref{Fig1}(a)), as suggested for Na$_3$Bi, Cd$_3$As$_2$ or ZrTe$_5$, where the field splits a Dirac point into two Weyl points of opposite chirality \cite{Cd3As2ChiralAnom,Cd3As2ChiralAnom2,Na3BiChiralAnom,ZrTe5Chiral}. In GdPtBi both the lack of inversion symmetry and an applied field appear to be important ingredients for their realization \cite{GdPtBiChiral}. A third route would be for the Weyl points to arise due to time reversal symmetry breaking from a magnetically ordered state (route III in Fig.~\ref{Fig1}(a)). This has been proposed to occur in YbMnBi$_2$ due to a canted antiferromagnetic state on the basis of angle-resolved photoemission spectroscopy (ARPES) \cite{YbMnBi2ARPES},  while the presence of Dirac fermions was suggested from magnetotransport measurements \cite{YbMnBi2Magneto}.

Meanwhile, understanding the role of topology in strongly correlated systems has also become a focus of attention due to the proposal for topological Kondo insulators, e.g. SmB$_6$, where surface states exist within the bulk gap opened by the hybridization between coherent $4f$ and conduction electrons \cite{SmB6a,SmB6f,SmB6c,SmB6d,SmB6b}. Similarly it is greatly desirable to look for systems with strong correlations which are gapless but topologically non-trivial, with Dirac or Weyl fermions. Recently, a heavy Weyl fermion state was theoretically proposed to occur in CeRu$_4$Sn$_6$ \cite{PhysRevX7011027}, which has a non-trivial band topology \cite{sundermann2015ceru4sn6}.  In order to search for the third route to Weyl fermions in a system with strong electronic correlations, we studied the magnetic Kondo compound CeSb, which is a cubic material with the rocksalt structure (Fig.~\ref{Fig1}(b)) that shows both a magnetically ordered ground state below $T_N~=16$~K, and evidence for strong electronic correlations. The  field-temperature phase diagram displays numerous magnetic phases, all of which consist of differently stacked arrangements of sheets of Ce atoms which are either paramagnetic, or order ferromagnetically with moments parallel or anti-parallel to the stacking direction \cite{CeSbMag}. Due to  an odd number of band inversions in the electronic structure, CeSb has also been proposed to have a non-trivial band topology \cite{CeSbarxiv}, as have various La based monopnictides \cite{zeng2015topological}. The presence of the Kondo interaction in CeSb is suggested from a logarithmic temperature dependence of the resistivity, and acoustic quantum oscillation measurements point to the presence of heavy electron bands, which arise due to hybridization between the conduction and $4f$ electron bands \cite{CeSbHeavyhole}. Here we report evidence for the emergence of Weyl fermions in the field-induced ferromagnetic state of CeSb, from the observation of a negative magnetoresistance when the current and applied field are parallel. The presence of Weyl points is also supported by electronic structure calculations, which indicate that the ferromagnetic order is important for their realization. These results suggest that CeSb may  represent a new class of materials with both Weyl fermions and strong electronic correlations.

\section*{Results}

\begin{figure}[h]
\begin{center}
  \includegraphics[width=0.75\columnwidth]{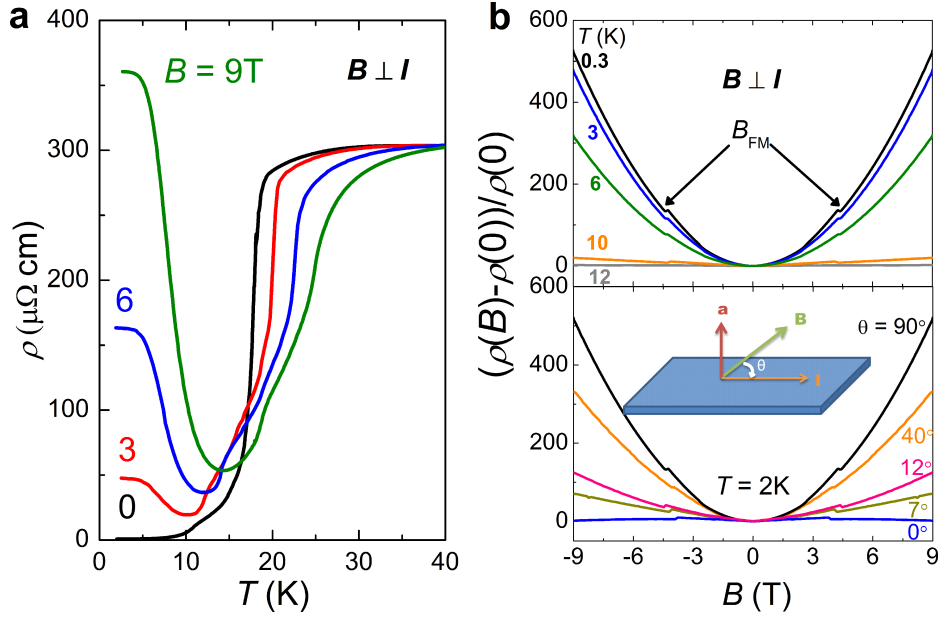}
\end{center}
	\caption{\textbf{Large positive magnetoresistance in CeSb for the magnetic field applied perpendicular to the current.} (a) Temperature dependence of the electrical resistivity of CeSb in various applied fields with $\boldsymbol{B}\perp\boldsymbol{I}$. At low temperatures the field leads to a significant enhancement of the resistivity indicating a large positive magnetoresistance. (b) Upper panel: Magnetoresistance as a function of applied field  with $\boldsymbol{B}\perp\boldsymbol{I}$ at various temperatures, demonstrating that the positive magnetoresistance becomes much more significant below 10~K. The arrows point to $B_{\rm{FM}}$, the field at which there is a transition from an AFF to the FM state with increasing field. (b) Lower panel: Magnetoresistance as a function of applied field for different $\theta$, where $\theta$  is the angle between the applied field and the current. As $\theta$ is reduced from $90^{\circ}$, the magnetoresistance decreases and becomes negative at high field near $0^{\circ}$.}
   \label{Fig2}
\end{figure}

\noindent\textbf{Sample characterization.} The magnetic susceptibility of single crystals of CeSb is displayed in  Fig.~\ref{Fig1}(c), which shows a sharp peak at around 17~K before dropping abruptly, due to the onset of magnetic order. The transition corresponds to entering the antiferromagnetic phase with paramagnetic layers (AFP), where there is no net magnetization and the zero-field cooling (ZFC) and field cooling (FC) curves do not split. At around 8~K  there is a significant difference between the ZFC and FC data which signals the onset of the antiferro-ferromagnetic (AFF) phase where all the Ce layers order ferromagnetically within the layer \cite{CeSbMag}. The electrical resistivity ($\rho(T)$)  of both CeSb and LaSb is shown in Fig.~\ref{Fig1}(d).  While LaSb behaves as a simple metal, $\rho(T)$ of CeSb is significantly enhanced. Upon cooling there is a decrease of $\rho(T)$  from 300~K to 80~K below which it increases reaching a maximum at 35~K. Figure~\ref{Fig1}(e) shows the magnetic contribution to $\rho(T)$ demonstrating that below 80~K there is a logarithmic increase of the magnetic resistivity, likely due to incoherent Kondo scattering, followed by a decrease below 35~K, presumably arising from the onset of coherence. 
\newline
\newline
\noindent\textbf{Angular-dependence of the magnetoresistance.} A particularly important signature of Weyl fermions  is  a negative magnetoresistance when the applied field is parallel to the current direction which arises due to a population imbalance between Weyl fermions of different chiralities induced by a magnetic field, resulting in a net current \cite{Nielsen1983,ChiralAnomTheor1}. We therefore measured the magnetoresistance of CeSb as a function of angle and temperature. Figure~\ref{Fig2}(a) displays $\rho(T)$ measured in various applied fields along [001], perpendicular to the current direction.  In zero field,  $\rho(T)$ continues to decrease down to the lowest temperature. Upon applying a magnetic field there is a minimum of the resistivity before it significantly increases at low temperatures, indicating a very large positive magnetoresistance, before flattening at the lowest temperatures. The magnetoresistance is shown in Fig.~\ref{Fig2}(b) at several low temperatures. Below 10~K, there is an anomaly around 4.3~T which indicates the transition from an AFF phase to a field-induced ferromagnetic (FM) state.  At higher temperatures in the magnetic state, the transition to the FM phase is pushed to higher fields which is consistent with previous reports \cite{CeSbMag}. While the data at 12 and 10~K also show a positive magnetoresistance, at 6~K the enhancement is significantly greater. The magnetoresistance continues to become stronger with decreasing temperature, namely by  a factor of $\approx520$ at 0.3~K and 9~T with no indication of saturation, similar to the isostructural non-magnetic compounds LaSb and  LaBi  \cite{LaSbCava,Tafti21062016,LaBiHCl}.   Such a large positive magnetoresistance when the current and field are perpendicular has  been found in materials proposed to display  a chiral anomaly \cite{WeylTaAsChirAnom1,Na3BiChiralAnom,ZrTe5Chiral,Cd3As2HighMR,NbPHighMR}.

\begin{figure}[h]
\begin{center}
  \includegraphics[width=0.75\columnwidth]{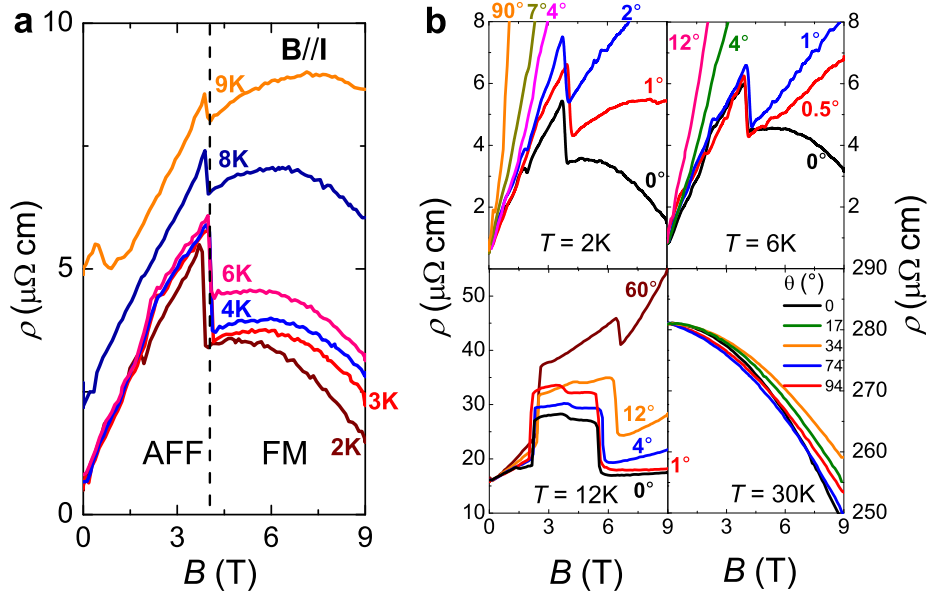}
\end{center}
	\caption{\textbf{Negative magnetoresistance in CeSb when the magnetic field is applied parallel to the current.} (a) Electrical resistivity as a function of  applied field with $\boldsymbol{B}\parallel\boldsymbol{I}$ at various temperatures up to 9~K, showing the onset of negative magnetoresistance in the ferromagnetic (FM) state. The dashed line marks the boundary between the antiferro-ferromagnetic (AFF) and FM states at around 4~T.  (b) Electrical resistivity as a function of applied field for different angles between the applied field and current. At 2~K and 6~K, the negative magnetoresistance is destroyed by slightly rotating the angle away from $0^{\circ}$. At 12~K no negative magnetoresistance is observed at any angle up to at least 9~T, while at 30~K negative behaviour is observed at all angles. }
   \label{Fig3}
\end{figure}

We also measured the magnetoresistance upon varying the angle $\theta$ between the current $\boldsymbol{I}$ and applied field $\boldsymbol{B}$.  As shown in  Fig.~\ref{Fig2}(c), at 2~K the magnetoresistance undergoes a significant decrease as $\theta$ is reduced, becoming negative near $0^{\circ}$. It can be seen more clearly in Fig.~\ref{Fig3}(a) that when the current and applied field are parallel, a negative magnetoresistance  appears above the transition from the AFF to the FM state.  As the temperature is increased, the decrease of the resistivity with field becomes less rapid and starts to be observed at  a higher field. At temperatures below 10~K, the negative magnetoresistance is very sensitive to the alignment of magnetic field and current and is destroyed by a slight deviation from  $0^{\circ}$, as shown by the  2~K and 6~K data in Fig.~\ref{Fig3}(b). Similar features were also observed on measurements of another sample where the current direction was rotated by an angle of about 20$^\circ$ compared to the one in  Fig.~\ref{Fig3}, indicating that the behaviour is reproducible when the current direction with respect to the crystal axes is changed (see Supplementary Information). 

In the AFF state, the magnetoresistance remains positive and shows a sharp drop at the transition to the FM state. This step-like reduction may be due to the sudden alignment of all the spins reducing spin disorder scattering. At 12~K the system goes through several field-induced magnetic phases, the so-called Devil's staircase \cite{CeSbMag}. In this case a weak positive magnetoresistance is observed in the FM state in the measured field range $6~$T$<B<9$~T. It may be that higher fields are required to realize the angle-sensitive negative magnetoresistance at 12~K, which appears only to onset in the FM state, at fields higher than that required to saturate the magnetization. In fact above 9~K where the magnetization shows additional field-induced magnetic transitions and larger applied fields are required to reach saturation (see Supplementary Information), we no longer observe the angle-sensitive negative magnetoresistance in our measured field range.  The behaviour below 10~K, where there is a negative longitudinal magnetoresistance which is sensitive to small changes in $\theta$, along with a large positive magnetoresistance when $\theta$ is increased, is very much like that found in other proposed  Weyl fermion systems \cite{ChiralAnomTheor1,WeylTaAsChirAnom1,WTe2ChiralAnom,NbPChiralAnom,WTe2ChiralAnom,GdPtBiChiral,Cd3As2ChiralAnom,Na3BiChiralAnom,ZrTe5Chiral,Cd3As2HighMR,NbPHighMR}. If the origin of this negative magnetoresistance were due to magnetic scattering, it would not be expected to disappear so rapidly with increasing $\theta$, nor would the decrease be expected to be so large as observed at low temperatures. The sharp drop in the magnetoresistance at the transition between the AFF and FM state does likely originate from the increasing alignment of spins with the applied field as a result of the change of magnetic structure, where the magnetization at 2~K shows a jump from 0.83$\mu_B$/Ce to 2.25$\mu_B$/Ce  (see Supplementary Information). However at higher fields the magnetization is almost entirely saturated, increasing by less than 1\% up to 10~T, while the overall magnitude of the resisitivity drop in the longitudinal magnetoresistance is similar to that going from the AFF state to FM state. Therefore it is not possible to account for the negative magnetoresistance at high fields from the increasing alignment of the spins in field.

\begin{figure}[h]
\begin{center}
  \includegraphics[width=0.99\columnwidth]{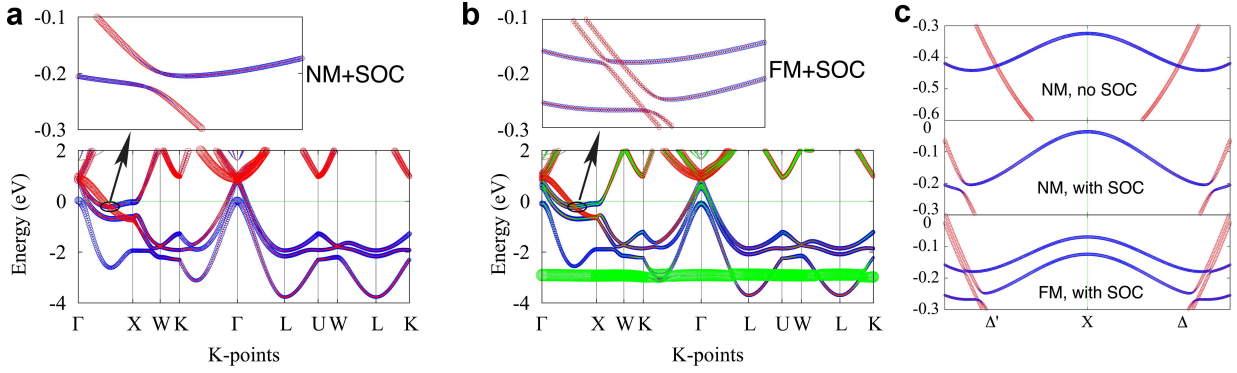}
\end{center}
	\caption{\textbf{Electronic structure calculations for CeSb.} (a) The electronic structure of CeSb in the non-magnetic case but including spin-orbit coupling. The red and blue colours represent Ce-$5d$ and Sb-$5p$ orbitals, respectively.  A magnification of the region where the band inversion occurs between $\Gamma$ and $\rm{X}$ is also displayed, which is fully gapped. (b) The electronic structure of CeSb in the ferromagnetic state. In addition to the Ce-$5d$ and Sb-$5p$ orbitals, the Ce-$4f$ orbitals are now explicitly included and are shown by green. The inset also shows an enlarged view of the band inversion, where two Weyl points between $\Gamma$ and $\rm{X}$ can be seen.(c) Diagram displaying how the combination of spin-orbit coupling (SOC) and ferromagnetism (FM) can lead to the emergence of two pairs of Weyl points. On the top the case is shown for the non-magnetic system without spin-orbit coupling where there are two band crossings with four-fold degeneracy at the crossing points. Turning on the spin-orbit coupling gaps out these points and generates  band inversions. In the ferromagnetic state, the Zeeman splitting again leads to gapless regions but with half the degeneracy of the first case, signifying the appearance of Weyl fermions.}
   \label{Fig4}
\end{figure}

 Current jetting has also been proposed to lead to negative magnetoresistance  \cite{Currentjet}. To minimize this effect, parallel platinum wires were spot welded across the whole  width of the sample face, with a large ratio of the length between the wires to the width (6.5 for the sample measured in Fig.~\ref{Fig3} and 15.7 for sample no.~2 in the Supplementary Information). In addition, we also measured the angle dependent magnetoresistance of the  isostructural non-magnetic reference compound LaSb (see Supplementary Information). Our measurements show a similar massive transverse magnetoresistance,  where the resistivity  in 9~T at 2~K is 430 times the zero field value. However, LaSb does not show a negative longitudinal magnetoresistance, despite  these measurements being prepared using the  same method as for CeSb, which is consistent with previous reports of LaSb \cite{LaSbCava,Tafti21062016}. Furthermore, the ratio of the transverse and longitudinal resistivities at 2~K are greater in LaSb at the field where the negative longitudinal magnetoresistance onsets in CeSb and therefore if current jetting were the cause of the negative magnetoresistance, it would also be expected to be seen in LaSb. As such this is not likely to be the origin of the behaviour in CeSb but nevertheless we cannot completely exclude  the influence of current jetting on the magnetoresistance. At higher temperatures, as displayed for 30~K, a negative magnetoresistance is observed at all angles, which is much weaker than the angle-sensitive results at lower temperature. This may arise due to the spins increasingly aligning in field and/or from the Kondo effect \cite{KondoMR}, and is unlikely to arise due to the chiral anomaly. 
\newline
\newline
\noindent\textbf{Theoretical calculations.} Since we observe possible evidence for the chiral anomaly at low temperatures in the FM phase, it is instructive to compare the band structures in the non-magnetic and FM cases, as displayed in Figs.~\ref{Fig4}(a) and (b) respectively. The electronic states near the Fermi level ($E_F$) are dominated by Ce-$5d$ (red) and Sb-$5p$ (blue) states. The effect of the spin-orbit coupling (SOC) in the non-magnetic case  is to generate two "band inversion" features in some of the electron bands in the regions $\Gamma~(0,0,0)-{\rm X}$ $(\pi,0,0)$, and $\Gamma-\bar{{\rm X}}$ $(-\pi,0,0)$  at around 0.2~eV below $E_F$ (see inset of Fig.~\ref{Fig4}(a)). This  is similar to what is typically observed in topological insulators, but with a very small splitting of around 20~meV. Moreover the calculated $Z_2$ classification is (1;000)(Table~\ref{Tab1}), supporting the previous findings of a non-trivial band structure \cite{CeSbarxiv}. For the calculation in the FM state shown in Fig.~\ref{Fig4}(b), the Ce $4f$ orbitals are explicitly included using GGA+U, and these form low lying localized states at around 3~eV below $E_F$, consistent with ARPES measurements \cite{CeSbarxiv}. The internal field in the FM state breaks time reversal symmetry, separating the bands in the region of the band inversion to form the new band-crossing features at $(\pm0.294\pi,0,0)$ and $(\pm0.308\pi,0,0)$ . We confirm that these features are indeed Weyl points by explicitly calculating the Berry curvature around them and the corresponding Chern number (see Supplementary Information). The energy dispersion close to these points can be fitted to the generalized Hamiltonian of Weyl fermions \cite{soluyanov2015type}  (see Supplementary Information). The detailed properties of these Weyl points depend on the choice of onsite Coulomb energy $U$, but their existence is robust within a reasonable range, from at least 5.0~eV to 9.0~eV.  Since the crystal has an inversion centre, the existence of Weyl points can also be inferred from the appearance of odd pairs of negative parity eigenvalues at eight time reversal invariant momenta (Table~\ref{Tab1}) \cite{Weyl1}. The emergence of Weyl points can be understood from the diagram shown in Fig.~\ref{Fig4}(c). In the presence of time reversal symmetry but no SOC there are four-fold degenerate band crossings, but when SOC is added and time reversal symmetry is broken, Weyl points are induced with half the degeneracy.
\section*{Discussion}
Both electronic structure calculations and magnetoresistance measurements give evidence for the emergence of Weyl points in the FM state of CeSb due to time reversal symmetry breaking. These results indicate the importance of the FM state for observing  Weyl fermions in CeSb, since the angle-sensitive negative magnetoresistance is only seen in the FM state once the magnetization has been saturated. Furthermore, the FM phase leads to a splitting of $\approx$40~meV of the Ce-$5d$ bands causing the Weyl points, which is estimated to be the equivalent of applying an external field of over 250~T. This case differs from antiferromagnetic GdPtBi  \cite{GdPtBiChiral}, where the chiral anomaly is insensitive to the magnetic ordering. Although Weyl fermions were suggested to be induced by the magnetic state of YbMnBi$_2$ from ARPES measurements \cite{YbMnBi2ARPES}, evidence for a chiral anomaly from a negative longitudinal magnetoresistance is not yet observed \cite{YbMnBi2Magneto}. Furthermore, enhanced effective electronic masses have been found experimentally in CeSb \cite{CeSbHeavyhole}, most strongly in the hole pocket, indicating the importance of electronic correlations  for the electronic structure. The presence of Weyl fermions in a low carrier density Kondo system CeSb would therefore provide a valuable opportunity for studying the interplay between topology, magnetism and electronic correlations.  However, further studies are necessary to look for additional evidence for Weyl fermions in CeSb. Although the requirement of an applied magnetic fields rules out the use of ARPES to detect them, evidence may be found from quasiparticle interference studies.
\section*{Methods} 
\noindent\textbf{Sample growth.} Single crystals of CeSb and LaSb were grown using a Sn flux with a molar ratio of Ce/La:Sb:Sn of 1:1:20 \cite{canfield1992growth}. The materials were sealed in an evacuated quartz tube, heated to 1150$^\circ$ before being cooled slowly to 800$^\circ$C. After centrifuging, the typical crystal size is $3\times3\times3$~mm. 
\newline
\newline
\noindent\textbf{Physical property measurements.} Resistivity measurements were performed using a Quantum Design Physical Property Measurement System (PPMS) from 300~K to 2~K either with a 14T-PPMS, or a 9T-PPMS with the sample rotation option. Resistivity measurements were performed after spot welding four Pt wires to the surface in the four probe geometry. When measuring the angle dependence of the magnetoresistance it was found that polishing the sample inevitably lead to a small misalignment, so the samples were instead cleaved on both sides. On one side the Pt wires were spot welded in a parallel alignment to reduce any possible effects of current jetting. The samples were in a regular rectangular geometry. For the sample described in the main text the thickness was 30.5~$\mu$m, the width parallel to the Pt wires was 101.7~$\mu$m and the length between the voltage wires was 651.5~$\mu$m. For the second sample shown in the Supplementary Information, the thickness was 41.1~$\mu$m, the width was 50.5~$\mu$m  and the length was 792.9~$\mu$m, while the LaSb sample on which the angle dependent magnetoresistance was measured  had a thickness of  140~$\mu$m, width of 377~$\mu$m and a length 1036~$\mu$m.   For measurements of the dc magnetization, a Quantum Design superconducting quantum interference device magnetometer and vibrating sample magnetometer option for the PPMS were both used.
\newline
\newline
\noindent\textbf{Electronic structure calculations.} Electronic structure calculations were calculated using density functional theory employing a plane-wave basis projected augmented wave method as implemented in the Vienna Abinitio Simulation Package (VASP). Whenever possible, we checked our results with the full potential LAPW method. We used  plane-wave basis up to 600~eV and a $12\times12\times12$ $\Gamma$ centered K-mesh. For the calculations in the non-magnetic state the Ce-$4f$ orbitals were treated as core electrons while in the magnetic state, these orbitals were taken into account using the GGA+U method with $U_f~=~6.7$~eV and $J_f~=~0.66$~eV. In all calculations spin-orbit coupling was treated using the second variational scheme.
\section*{Acknowledgments} We would like to thank Yi Zhou, Xin Lu, Yang Liu and Hanoh Lee for valuable discussions. This work was supported by the National Key R\&D Program of China (No.~2017YFA0303100, No.~2016YFA0300202, No.~2014CB648400), the National Natural Science Foundation of China (No.~U1632275, No.~11474251, No.~11274006) and the Science Challenge Program of China.
\section*{Additional information} Correspondence and requests for materials should be addressed to H. Q. Yuan (hqyuan@zju.edu.cn)
\section*{Author contributions} The project was concieved by C.Y.G. and H.Q.Y.. The crystals were grown by C.Y.G., F.W. and Y.J.Z.  C.Y.G. and  F.W. performed the measurements, which were analyzed by C.Y.G., M.S, F.W, F.S. and H.Q.Y. Electronic structure calculations and theoretical analysis were performed by C.C., F.W. and F.C.Z.. The manuscript were written by C.Y.G., C.C., M.S. and H.Q.Y with input from F.C.Z. and F.S.. All authors participated in discussions. 

\section*{Competing financial interests} 
The authors declare no competing financial interests.

\newpage

\begin{table}[h]
  \caption{Band parities at time reversal invariant momenta (TRIMs) in the non-magnetic (NM) and ferromagnetic  (FM) states. We list only states above $E_F$-4.0 eV, which are well separated from lower energy states,  ordered from lower energies to higher energies. The states on different sides of the single vertical lines are well separated by the SOC splitting, i.e. the states on the left (right) side of the dashed line are fully occupied (unoccupied) at $L$, respectively.  In the FM state, the lower partially occupied states separated by the SOC splitting have nine pairs of negative parity eigenvalues, demonstrating the existence of Weyl points in the system. In addition, due to the alignment of the magnetic moments, the three-fold rotational symmetry of the crystal structure is broken. The band-crossing features along the directions perpendicular to the moment direction will therefore form nodal lines instead of Weyl points, leading to only two pairs of Weyl nodes in the moment direction.}
  \begin{tabular}{c||c|c||c|c}
    TRIM & \multicolumn{2}{c||}{NM} & \multicolumn{2}{c}{FM} \\
   \hline
   $\Gamma$ (0, 0, 0)         &  $- + +$ & $- -$ & $- - - - - - -$ & $+ + + +$ \\
   $L'$ ($\pi$, 0, 0)             & $+ + +$ & $+ +$ & $+ + - + + + +$ & $- - - +$ \\
   $X$ ($\pi$, $\pi$, 0)        & $- + -$ & $- +$ & $- - - + - - +$ & $- - + +$ \\
   $L$  ($\pi$, $\pi$, $\pi$) & $+ + +$ & $+ +$ & $+ + - + + + +$ & $- - - +$  \\
  \end{tabular}
   \label{Tab1}
\end{table}

\end{document}